\documentclass{emulateapj}
\usepackage[pdftex]{graphicx}
\usepackage{amsmath}
\usepackage{color}

\slugcomment{Draft version}

\newcommand{\hmsun}{h^{-1}{\rm M}_\odot}
\newcommand{\hmpc}{h^{-1}{\rm Mpc}}

\newcommand{\msun}{{\rm M}_\odot}

\newcommand{\kms}{{\rm ~km~s^{-1}}}
\newcommand{\sag}{\texttt{SAG}}

\shorttitle{SAMs calibration using PSO}
\shortauthors{Andr\'es N. Ruiz et al.}

\begin{document}

\title{Calibration of semi-analytic models of galaxy formation using Particle Swarm Optimization}

\author{Andr\'es N. Ruiz\altaffilmark{$\dagger$,1,2,3}, 
	Sof\'ia A. Cora\altaffilmark{3,4,5},
	Nelson D. Padilla\altaffilmark{6,7}, 
	Mariano J. Dom\'inguez\altaffilmark{1,2,3},
	Cristian A. Vega-Mart\'inez\altaffilmark{3,4},
        Tom\'as E. Tecce\altaffilmark{6,7}, 
	\'Alvaro Orsi\altaffilmark{6,7}, 
	Yamila Yaryura\altaffilmark{1,2}, 
	Diego Garc\'ia Lambas\altaffilmark{1,2,3}, 
	Ignacio D. Gargiulo\altaffilmark{3,4,5}, 
        Alejandra M. Mu\~noz Arancibia\altaffilmark{6}}

\altaffiltext{$\dagger$}{andresnicolas@oac.uncor.edu}
\altaffiltext{1}{Instituto de Astronom\'{\i}a Te\'orica y Experimental, CONICET-UNC, Laprida 854, X5000BGR, C\'ordoba, Argentina.}
\altaffiltext{2}{Observatorio Astron\'omico de C\'ordoba, Universidad Nacional de C\'ordoba, Laprida 854, X5000BGR, C\'ordoba, Argentina.}
\altaffiltext{3}{Consejo Nacional de Investigaciones Cient\'{\i}ficas y T\'ecnicas, Rivadavia 1917, C1033AAJ Buenos Aires, Argentina.}
\altaffiltext{4}{Instituto de Astrof\'isica de La Plata, CONICET-UNLP, Paseo del Bosque s/n, B1900FWA, La Plata, Argentina.}
\altaffiltext{5}{Facultad de Ciencias Astron\'omicas y Geof\'{\i}sicas, Universidad Nacional de La Plata, Paseo del Bosque s/n, B1900FWA, La Plata, Argentina.}
\altaffiltext{6}{Instituto de Astrof\'{\i}sica, Pontificia Universidad Cat\'olica de Chile, Av. Vicu\~na Mackenna 4860, Santiago, Chile.}
\altaffiltext{7}{Centro de Astro-Ingenier\'{\i}a, Pontificia Universidad Cat\'olica de Chile, Av. Vicu\~na Mackenna 4860, Santiago, Chile.}

\begin{abstract}
We present a fast and accurate method to select an optimal set of parameters in
semi-analytic models of galaxy formation and evolution (SAMs). Our approach
compares the results of a model against a set of observables applying a
stochastic technique called Particle Swarm Optimization (PSO), a self-learning
algorithm for localizing regions of maximum likelihood in multidimensional
spaces that outperforms traditional sampling methods in terms of computational
cost. We apply the PSO technique to the $\sag$ semi-analytic model combined
with merger trees extracted from a standard $\Lambda$CDM {\em N}-body
simulation.  The calibration is performed using a combination of observed
galaxy properties as constraints, including the local stellar mass function and
the black hole to bulge mass relation. We test the ability of the PSO algorithm
to find the best set of free parameters of the model by comparing the results
with those obtained using a MCMC exploration. Both methods find the same
maximum likelihood region, however the PSO method requires one order of
magnitude less evaluations. This new approach allows a fast estimation of the
best-fitting parameter set in multidimensional spaces, providing a practical
tool to test the consequences of including other astrophysical processes in
SAMs.
\end{abstract}



\section{Introduction}
\label{sec:introduction}

The combination of numerical methods and computational resources provides one
of the best tools to study the origin and evolution of structures in the
Universe. The Lambda Cold Dark Matter ($\Lambda$CDM) scenario for the growth of
dark matter (DM) structures has been extensively studied using numerical {\em
N}-body simulations, which follow the evolution of such structures from shortly
after decoupling to the present-day over a large dynamical range 
\citep{springel_millennium_2005,boylan-kolchin_millenniumII_2009,
klypin_bolshoi_2011,angulo_millenniumXXL_2012}.
 
Studying the evolution of baryonic matter is more complex due to the highly
non-linear effects of gas, making it impossible (with the current
state-of-the-art) to perform a fully self-consistent numerical simulation that
resolves all the scales relevant to the physics of galaxy formation in a large
volume. In addition, physical processes in the sub-grid scale, such as the
star-formation process and different feedback mechanisms are still poorly
understood due to their intrinsic complexity. This is why semi-analytic models
of galaxy formation (SAMs) play a fundamental role, since they can produce
large samples of galaxies at a low computational cost which can be used to
study properties such as star formation, luminosity, colors and chemical
evolution by following the evolution of baryons in a simplified way
\citep{baugh_review_2006,benson_review_2010,silk_review_2012,
frenk_review_2012}.

A downside of using SAMs is that they depend on a large number of free
parameters which are included in the implementations of the different physical
processes modeled; these parameters are usually calibrated by choosing a set of
observables that the simulated galaxies are expected to reproduce.
\cite{kampakoglou_mcmc_2008} and \cite{henriques_mcmc_2009} introduced the
Monte Carlo Markov Chains (MCMC) technique to carry out a statistical
exploration of the multidimensional parameter space of a SAM. Since this
technique is well known, MCMC quickly became a popular tool for calibrating the
free parameters of SAMs, using as observational constraints not only $z=0$ data
\citep{henriques_mcmc_2009, henriques_mcmc_2010,lu_bayes_2011,lu_kband_2012}
but also observed or inferred properties at higher redshifts
\citep{mutch_sam_2013,mutch_sam2_2013,henriques_mcmc_2013,henriques_mcmc_2014}.

More refined Bayesian techniques for determining the value of the free
parameters in SAMs were introduced by \citet{bower_parameter_2010} with the
Gaussian model emulator technique (ME) (see e.g. \citealt{kennedy_me_2001}), an
approach based in building a statistical predictor for the results from a given
model based on a limited set of model runs. The ME technique was also
implemented by \cite{gomez_parameter_2012,gomez_parameter_2014} to study the
impact of formation histories on the final satellite population of Milky
Way-sized galaxies.

The increasing complexity of SAMs motivate the search for computationally more
effective and fast methods to perform calibrations. In this work we apply for
the first time another machine learning inspired sampling method, called
Particle Swarm Optimization (PSO, \citealt{kennedy_pso_1995}) to select the
optimal set of parameters for a SAM. In high-dimensional spaces, such as the
SAM parameter space analysed here, the PSO technique can outperform the
traditional methods in terms of computational cost in the search for an
adequate solution. Such increased search performance was reported by
\cite{prasad_pso_2012} who applied this technique for cosmological parameter
estimations using the WMAP7 cosmic background data. The aim of this paper is to
introduce the PSO technique as a fast calibrator of SAMs. A detailed comparison
of the performance between PSO and other techniques such as MCMC is studied in
\cite{prasad_pso_2012}. In this work, we further explore
this issue demonstrating that the PSO technique is very competitive in
comparison to MCMC, reaching much more rapidly a physically equivalent
convergence region. The plan of this paper is as follows. In Sec.
\ref{sec:sim}, we introduce the details of the  semi-analytic model used,
$\sag$ \citep[acronym for Semi-Analytic
Galaxies,][]{cora_sam_2006,lagos_sam_2008,padilla_flips_2014,gargiulo_afe_2014},
specifying the relevant free parameters to be explored. We also present the DM
{\em N}-body simulation used to construct the merger trees which are fed into
$\sag$ to construct the galaxy population. In Sec. \ref{sec:method}, we
describe the PSO implementation used to find the best set of $\sag$ parameters
and the methodology used to estimate the corresponding errors. The
observational constraints and likelihood function used to perform the
calibration are described in Sec. \ref{sec:obs}. In Sec. \ref{sec:results} we
present and discuss the results of a calibration, and the
suitability of the PSO approach for SAM parameter estimations,
{demonstrating its advantage over MCMC. Finally, in Sec. \ref{sec:conclusions}
we summarise the main results of this work.

%
%
\section{The simulated galaxy population} 
\label{sec:sim}

The galaxy populations used in this work were generated using a semi-analytic
model of galaxy formation that computes the evolution of the baryonic component
taking as input DM halo merger trees extracted from a numerical {\em N}-body
simulation. In this section we describe this procedure.


\subsection{N-body simulation}

We use a DM-only {\em N}-body simulation in the standard $\Lambda$CDM scenario,
run with \texttt{Gadget-2} \citep{springel_gadget2_2005} using $640^3$ particles
in a cubic box of comoving sidelength $L=150\hmpc$. For the cosmological
parameters we adopt the values $\Omega_{\rm m}=0.28$, $\Omega_{\rm b}=0.046$,
$\Omega_{\Lambda}=0.72$, $h=0.7$, $n=0.96$, $\sigma_{8}=0.82$, corresponding to
the WMAP7 cosmology \citep{jarosik_wmap7_2011}. The mass of a dark matter
particle is $m_{\rm dm}=1\times 10^9\,\hmsun$. The initial conditions were
generated using \texttt{GRAFIC2} \citep{bertschinger_grafic2_2001}. The
simulation was evolved from $z=61.2$ to the present epoch, storing 100 outputs
equally spaced in $\log_{10}(a)$ between $z=20$ and $z=0$
\citep{benson_trees_2012}.  

DM halos were identified in the simulation outputs using a friends-of-friends
(FoF) algorithm, and then self-bound substructures (subhalos) are extracted
using \texttt{SUBFIND} \citep{springel_subfind_2001}. We consider only
(sub)halos with at least 10 particles. The position and velocity of the
most-bound particle in each subhalo is stored, as these are used by the SAM to
trace the positions and velocities of galaxies within FoF halos, assuming that
galaxies trace the DM distribution.

With the aim of speeding-up the calibration process, we follow a scheme similar
to that implemented by \citep{henriques_mcmc_2009,henriques_mcmc_2013}.  We
split the simulation into $64$ sub-boxes that contain complete merger-trees.
From these $64$ sub-boxes, we select $4$ that sample different simulation
environments to perform the calibration (with both PSO and MCMC methods). Once
the model is calibrated, we run $\sag$ using all the merger-trees of the
simulation. We do not find any bias when comparing the results obtained from
the sample of $4$ sub-boxes with those of the whole simulation.

%
\subsection{Semi-analytic model of galaxy formation $\sag$} 
\label{sub:sag}

We use the semi-analytic model $\sag$ based on the Munich semi-analytic model
described by \citet{springel_subfind_2001} and further modified and developed
by \cite{cora_sam_2006}, \cite{lagos_sam_2008}, \cite{padilla_flips_2014} and
\cite{gargiulo_afe_2014}.

Subhalo merger trees are used as inputs for $\sag$; galaxies are assumed
to form in the centre of DM subhalos. A fraction of the hot gas in the halo
loses energy via radiative cooling and settles in the centre, forming a gaseous
disc with an exponential density profile. Star formation begins when the
density of this cold gas disc becomes high enough. The most massive subhalo
within a FoF halo hosts the central galaxy, which we call a {\it type 0}
galaxy. Cold gas in galaxies of this class can be replenished by infall of
cooling gas from the intergalactic medium. 

Galaxies in smaller subhalos of the same FoF halo are considered as satellites
and labeled as {\it type 1}. If the satellites subhalos are not longer
identified by the {\tt SUBFIND} halo finder, then the galaxy contained within
are not assumed to be destroyed but preserved until eventually merges with
the central galaxy of its host subhalo after a dynamical friction time-scale.
These galaxies are labeled as {\it type 2} satellites. 

When a galaxy becomes a satellite of either type, all of its hot gas halo is
removed and transferred to the hot gas component of the corresponding type~0
galaxy; consequently, gas cooling is suppressed in all satellites. Only
galaxies of type~0 and~1 can continue accreting stars and gas from merging
satellites. In major mergers, characterised by a satellite
with baryonic mass larger than $30$ per cent that of the central
galaxy, the stellar disc of the remnant is transfered to the bulge,
while in minor mergers only the stars of the merging satellite are transferred
to the bulge component of the central galaxy. Mergers and disc instabilities
trigger starbursts which contribute to the formation of a bulge component.
These starbursts are characterised by a certain time-scale in
which the cold gas that has been transfered to the bulge is gradually
consumed \citep{gargiulo_afe_2014}. We consider the disc instability
criterion to evaluate the triggering of bursts in the remnant galaxy
after a merger. This condition is inspired in the fact that the
material accreted onto the galactic disc has misaligned angular momenta
\citep{padilla_flips_2014}. The flips in angular momenta seen in the
dark matter haloes are assumed to be the same for the cold baryons, but
we assume a loss of specific angular momentum of the disc such that it
entails an inhibition of growth of the disc angular momentum due to
slews and flips. As a result of the drop in their dimensionless spin
parameter, galactic discs become slightly smaller, which impacts 
the quiescent star formation rate in the discs and the frequency of disc
instability events.

Star formation is regulated by energetic feedback from supernova (SNe)
explosions and active galactic nuclei (AGN) as implemented by
\citet{lagos_sam_2008}. The former induces transfer of gas and metals from the
cold to the hot gas phase, while the latter, which are a consequence of the
growth of supermassive black holes in galaxy centers, suppress gas cooling in
central galaxies. For this work we consider an scheme where the material
expelled from the galactic disc by SNe is ejected to an external reservoir and
reincorporated later to the hot gas phase (``ejection'' scheme). This is
supported by observed dependence of the baryon fraction on halo virial mass
\citep{delucia_sfr_2004}. The mass reincorporated in our model includes a
dependence on the virial velocity as in \citet{guo_sam_2011}.

The recycling process as a result of stellar mass loss and SNe explosions
contributes to the chemical enrichment of the different baryonic components,
as is described in detail in \citet{cora_sam_2006}. In the
current version of the model, we adopt the yields of
\citet{karakas_yields_2010} to follow the production of chemical elements
generated by low- and intermediate-mass stars (mass interval $1-8 \msun$).
Yields resulting from mass loss of  pre-supernova stars and explosive
nucleosynthesis are taken from \citet{hirschi_yields_2005}
and \citet{Kobayashi2006}, respectively. This combination of stellar yields are
in accordance with the large number of constraints for the Milky Way
\citep{romano_yields_2010}. For the ejecta from type Ia SNe (SNe Ia), we
consider the nucleosynthesis prescriptions from the updated model W7 by
\citet{Iwamoto99}.  The SNe Ia rates are estimated using the single degenerate
model \citep{Greggio83, Lia2002}.  This model involves the fraction of binary
systems whose components have masses between $0.8$ and $8$M$_{\odot}$ which are
progenitors of SNe Ia, for which we adopt a value of $0.04$ that gives a rate
which evolves with redshift in good agreement with the compilation of data
given by \cite{melinder_sne_2012}. The return time-scale of mass losses and
ejecta from all sources considered are estimated using the Stellar lifetime
given by \citet{Padovani93}. Stellar winds from low and intermediate mass
stars, core-collapse SNe and SNe Ia contribute to the metal enrichment of the
cold gas from which different generations of stars form.  Each star forming
event is characterised by a Chabrier initial mass function (IMF)
\citep{chabrier_2003}. The chemical contamination of the hot gas through SNe
feedback also affects the metal-dependent gas cooling rate, estimated by
considering the total radiated power per chemical element given by
\cite{foster_xray_2012}.

This version of $\sag$ has several free parameters that have to be tuned to
reproduce observational data.  It allows to test the ability of the PSO
algorithm presented here to find the best set of parameters. 


\subsection{Free parameters of $\sag$} 
\label{sub:freeparam}

We choose to explore the behaviour of the $\sag$ model by tuning seven 
free parameters keeping the rest fixed at a given value. We describe here the role played 
by these seven free parameters. They are:\\

{\bf (i) $\alpha$ - star formation efficiency.} This parameter is involved in
the star formation process, which is implemented following
\cite{croton_cooling_2006}.  A galaxy forms stars only when its cold gas
exceeds a critical mass $M_{\rm cold,crit}$, with a rate as follow:
\begin{equation} {{dM_\star}\over{dt}} = \alpha {{M_{\rm cold} - M_{\rm
cold,crit}}\over{t_{\rm dyn}}}, \end{equation}
with
\begin{equation} M_{\rm cold,crit} = 3.8\times10^9 \left({V_{\rm
vir}}\over{200\kms}\right) \left({3R_{\rm disc}} \over{\rm~10kpc}\right) \msun,
\label{eq:sf_mcrit} \end{equation}
where $t_{\rm dyn}=V_{\rm vir}/3R_{\rm disc}$ is the dynamical time of the
galaxy, $V_{\rm vir}$ is the circular velocity at the virial radius and $R_{\rm
disc}$ the disc scale length given by $R_{\rm disc}=(\lambda/\sqrt{2})\,R_{\rm
vir}$, being $\lambda$ the spin parameter of the host halo
\citep{MoMaoWhite_1998}.\\

{\bf (ii) $\epsilon_{\rm disc}$ - SNe feedback efficiency associated
to the star formation (SF) taking place in the disc.} This controls the amount of cold
gas reheated by the energy released by SNe generated
from quiescent SF that occurred in the disc. The reheated mass produced by a
star forming event which generates a stellar mass $\Delta M_\star$ is assumed
to be
\begin{equation}
\Delta M_{\rm reheated} = {4 \over 3} \epsilon_{\rm disc} {{\eta E}\over{V_{\rm vir}^2}} 
\Delta M_{\star},
\label{eq:mreheated}
\end{equation}
where $E=10^{51}$ erg s$^{-1}$ is the energy generated by each supernova,
and $\eta$ is the number of SNe per each solar mass of stars
formed, computed from the assumed IMF normalised
between 0.1 and 100 $\msun$.\\

{\bf (iii) $\epsilon_{\rm bulge}$ - SNe feedback efficiency associated to the
SF taking place in the bulge.} This controls the amount of bulge cold gas
reheated by SNe formed in the bulge when a starburst is triggered.  Since the
cold gas transfered from the disc to the bulge is gradually consumed, it can
also be affected by SNe feedback. The reheated mass is also given by Eq.
(\ref{eq:mreheated}), but with the efficiency $\epsilon_{\rm disc}$ replaced by
$\epsilon_{\rm bulge}$.\\

{\bf (iv) $f_{\rm BH}$ - fraction of cold gas accreted onto the central
supermassive black hole (SMBH).} A SMBH grows via gas flows to the galactic core
triggered by the perturbations to the gaseous disc which result from galaxy
mergers or disc instabilities.  When a merger occurs, central SMBHs are assumed
to merge instantaneously. The mass of cold gas accreted by the resulting SMBH
is given by 
\begin{equation} 
\Delta M_{\rm BH} = f_{\rm BH} {{M_{\rm sat}}\over{M_{\rm cen}}} {{M_{\rm cold,sat} 
+ M_{\rm cold,cen}} \over {(1 + 280 \kms / V_{\rm vir})^2}},
\end{equation}
where $M_{\rm cen}$ and $M_{\rm sat}$ are the masses of the merging central and
satellite galaxies, and $M_{\rm cold, cen}$ and $M_{\rm cold,sat}$ are their
corresponding cold gas masses.  In the case of disc instabilities, only the
host galaxy is involved.\\

{\bf (v) $\kappa_{\rm AGN}$ - efficiency of cold gas accretion onto the SMBH
during gas cooling.} The cold gas accretion during gas cooling occurs once a
static hot gas halo has formed around the central galaxy, and is assumed to be
continuous. It is given by 
\begin{equation} 
{{dM_{\rm BH}}\over{dt}} = \kappa_{\rm AGN} {{M_{\rm BH}} \over{10^8 \msun}} 
{{f_{\rm hot}}\over{0.1}} \left({{V_{\rm vir}}\over{200 \kms}}\right)^2,
\label{eq:BHVvir2} 
\end{equation}
where $f_{\rm hot}=M_{\rm hot}/M_{\rm vir}$, being $M_{\rm hot}$ and 
$M_{\rm vir}$
the hot gas and virial masses, respectively\footnote{We now consider the
mass accretion rate to depend on the square of the virial velocity, instead of
on the cube of the velocity as in \cite{lagos_sam_2008}. The old prescription
caused the SMBHs at the centre of cluster-dominant galaxies to grow
unrealistically large, at the expense of the intracluster medium. With this
change, the accretion is consistent with a Bondi-type accretion
\citep{bondi_bh_1952}.}.\\

{\bf (vi) $D_{\rm pert}$ - factor involved in the distance scale of
perturbation to trigger disc instability}. 
A galactic disc that becomes unstable and is also perturbed by a neighbouring
galaxy will undergo a starburst; stars created in this event
contribute to the bulge formation. The stellar disc is also transfered to the
bulge. The stability to bar formation is lost when
\begin{equation} 
\frac{V_{\rm disc} }{ (G M_{\rm disc}/R_{\rm disc})^{1/2}} \le \epsilon_{\rm thresh}, 
\label{eq:diskinstab}
\end{equation}
where $M_{\rm disc}$ is the mass of the disc (cold gas plus
stars), and $V_{\rm disc}$ is the maximum circular velocity of the disc.  In
our model we adopt a value of  $\epsilon_{\rm thresh}=1$, in agreement with the
theoretical motivation for Eq. (\ref{eq:diskinstab}) \citep{Efstathiou+1982}.
We assume a galaxy to suffer the effects of the interaction when the mean
distance between galaxies sharing the same dark matter halo is smaller
than $D_{\rm pert}$ times the disc scale length of the unstable galaxy. The
effect of disc instability in the calibration process is regulated by $D_{\rm
pert}$.\\

{\bf (vii) $f_{\rm reinc}$ - fraction of ejected reheated cold gas that is
reincorporated into the hot halo gas.} In the ejection scheme used here, the
cold gas reheated by SNe explosions is expelled from the galactic disc and
stored in an external reservoir.  This material that leaves the halo is
reincorporated into the hot halo gas on a timescale which depends on the virial
velocity of the host halo. It is given by  
\begin{equation} {{dM_{\rm reinc}}\over{dt}} = f_{\rm reinc} {{M_{\rm ejec}}
\over{t_{\rm dyn,h}}} \left({{V_{\rm vir}}\over{220 \kms}}\right),
\label{eq:Mreinc} \end{equation} 
where ${M_{\rm ejec}}$ is the mass of the reheated cold gas that is ejected,
and $t_{\rm dyn,h}=R_{\rm vir}/V_{\rm vir}$ is the dynamical time of the halo.
We introduce the factor that involves the virial velocity following
\citet{guo_sam_2011}, thus taking into account the fact that the mass ejected
by lower mass systems is likely more difficult to be re-accreted since the wind
velocities are higher relative to the escape velocity.\\


\section{The Particle Swarm Optimization method}
\label{sec:method}

\subsection{The algorithm}

PSO is a computational technique originally introduced by
\cite{kennedy_pso_1995} to optimise multidimensional parameter explorations. If
$X=\{x_1,x_2,...,x_{\rm D}\}$ is a point in $\mathbb{R}^{\rm D}$ and $F(X)$ is
a fitness (or optimization) function which is a measure of the `quality' of
point $X$, the PSO algorithm computes $F(X)$ simultaneously at different points
in the multidimensional space using a set of `particles' which share
information, thus determining new exploratory positions from both their
individual and collective knowledge.  This process can be visualised as a {\it
swarm} of particles exploring iteratively the multidimensional space,
exchanging information as they do so. In the following, we introduce some
definitions in order to describe the PSO implementation used in this work.\\

{\bf Particles:} the `computational agents' which explore the
multidimensional space.  Each particle has an identification number
$i=1,...,N_{\rm p}$, where $N_{\rm p}$ is a free parameter of the PSO
algorithm. At every step $t$ the particles have a `position' $X_i(t)$ and a
`velocity' $V_i(t)$ in the multidimensional space.\\

{\bf Fitness function $F(X)$:} a function which evaluates the `quality' of a
point $X$, that is, the likelihood of the model reproducing a particular
constraint using that set of values as input. In this work we simply use a
$\chi^2$~statistic (described in more detail in Sec. \ref{sec:obs}).\\

{\bf Best individual value:} if $F_i^{\rm max}(t)$ is the best value of
$F(X)$ found by the $i$-th particle at step $t$, we label as $B_i(t)$ the
position $(X)$ of that point in the multidimensional parameter space, so that
\begin{equation} 
F_i^{\rm max}(t) = F[B_i(t)] \ge F[X_i(\tau)];~~\tau \le t.
\end{equation}

{\bf Best global value:} if we define $F^{\rm max}(t) = {\rm max}\{F_i^{\rm
max}(t)\}_ {i=1,...,N_{\rm p}}$ as the best value of $F(X)$ found for all the
particles at step $t$, the position of that point is labelled as $G(t)$ and
satisfies
\begin{equation}
F^{\rm max}(t) = F[G(t)] \ge F[B_i(t)];~~i=1,...,N_{\rm p}.
\label{eq_global_value}
\end{equation}

\subsubsection{Dynamics}
In a new time step, particle positions are updated following 
\begin{equation}
  X_i(t+1) = X_i(t) + V_i(t+1),
  \label{eq:pos}
\end{equation}
and the velocity is computed according to
\begin{equation}
  V_i(t+1) = wV_i(t) + c_1 \xi_1 [B_i(t) - X_i(t)] + c_2 \xi_2 [G(t) - X_i(t)].
  \label{eq:vel}
\end{equation}
The coefficient $w$ is the so-called {\it inertial weight}, $c_1$ and $c_2$ are
the {\it acceleration constants} which determine the contribution to the
velocity due to individual and collective learning, respectively, and $\xi_1$
and $\xi_2$ are random numbers drawn from a uniform distribution between $0$
and $1$.

The first term on the right hand side of Eq. (\ref{eq:vel}) moves the particle
along a straight line, whereas the second and third terms accelerate it toward
positions $B_i(t)$ and $G(t)$. There are several different implementations of
the PSO algorithm for astrophysical problems with different choices for the
parameters $w$, $c_1$ and $c_2$, or even with extra parameters
\citep{skokos_pso_2005,wang_pso_2010,rogers_pso_2011}. For this work, we choose
the set of parameters used by \cite{prasad_pso_2012}, which are the values
suggested in the PSO Standard 2006\footnote{{\tt
www.particleswarm.info/Standard\_PSO\_2006.c}} ({\tt PSOS06}), that is 
\begin{equation}
  w={{1}\over{2~{\rm ln(2)}}} \approx 0.72,
\end{equation}
\begin{equation}
  c_1=c_2=0.5+{\rm ln(2)} \approx 1.193.
\end{equation}
Adopting different values for these parameters do not change or improve  the
results of the explorations, as was also noted by \cite{prasad_pso_2012}; these
parameters have impact mainly on convergence times.  The standard values
adopted allow us to obtain fast and accurate results.

\subsubsection{Initial conditions}
Usually the initial positions and velocities of the particles are
assigned randomly, according to
\begin{equation}
  X_i(t=0) = X_{\rm min} + \xi \left(X_{\rm max} - X_{\rm min} \right),
\end{equation}
\begin{equation}
  V_i(t=0) = \xi V_{\rm max},
  \label{eq:ic_vel}
\end{equation}
where $\xi$ is a uniform random number between $0$ and $1$ and
$[X_{\rm min},X_{\rm max}]$ are the limits of the search space. However, in
this work we generate the initial positions of the particles using a Maximin
Latin Hypercube (MLH) \citep{stein_lh_1987}.  The MLH is an extension of the
traditional Latin Hypercube (LH) \citep{mckay_lh_1979}, a technique that
samples a multidimensional space more efficiently than a random distribution.
To construct a LH of $N_{\rm p}$ points, the search range of each parameter
must be divided into $N_{\rm p}$ equal parts; the points are then randomly
selected so that two points do not occupy the same interval for each of the
parameters.  A MLH run consists in generating thousands of LHs and selecting
the one with the greatest distance between any two points. We
use a MLH instead of a random distribution to initialize the position
of the particles motivated by the adoption of a small number of
particles to sample a multidimensional space and to avoid repeated
sampling of given regions. By construction, the MLH ensures that all
particles are well spatially separated without clustering and allowing
for best initial guess of the behaviour (quan- tified, in our case, by the
fitness function used) of the whole parameter space. An example for a
2-dimensional parameter space is shown in Fig.~\ref{fig:lh}. For velocities, we
adopt a random distribution as given by Eq. \ref{eq:ic_vel}.
\begin{figure}
\centering
\includegraphics[scale=0.28]{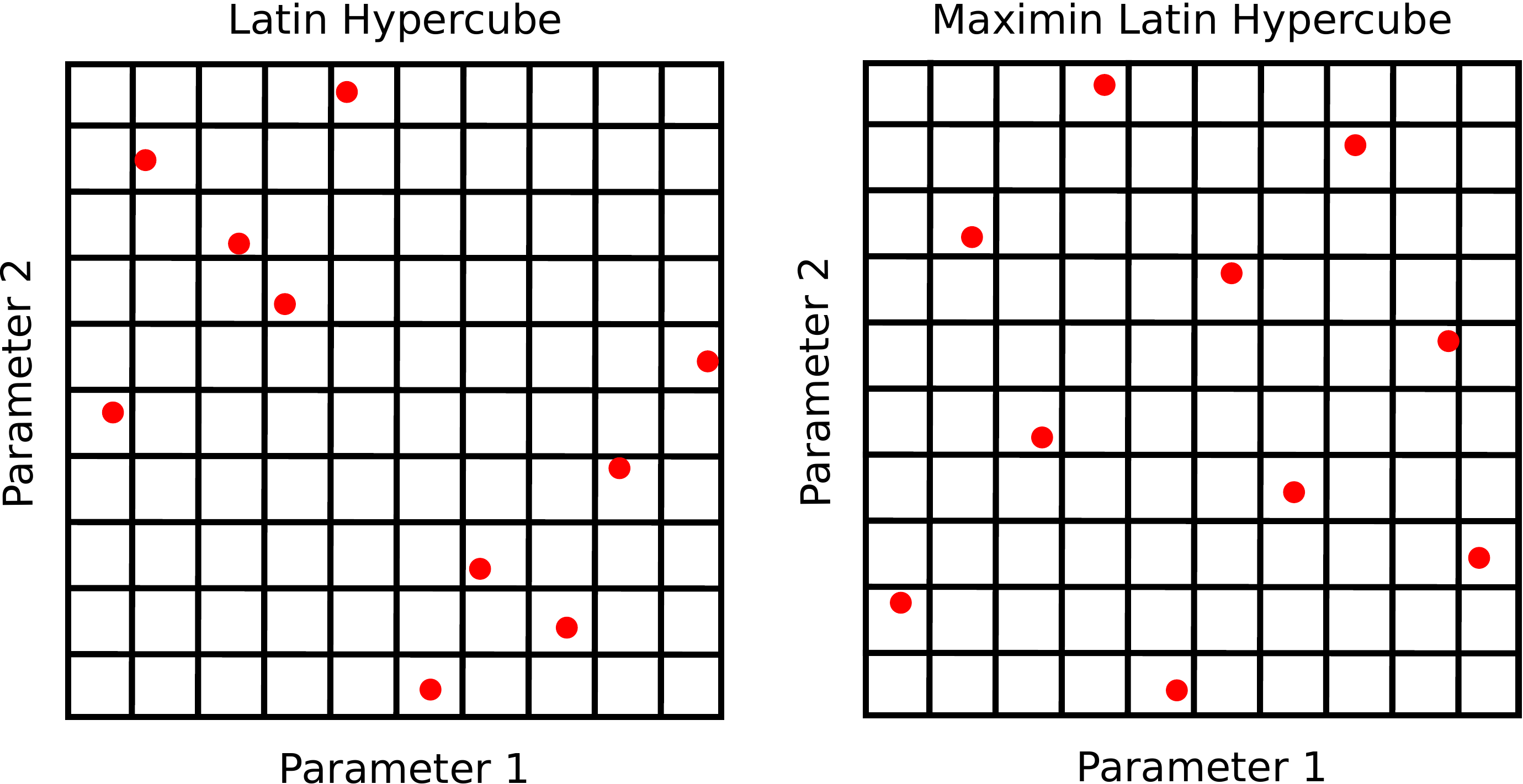}
\caption{Examples of a Latin Hypercube (LH) (left) and a Maximin Latin
Hypercube (MLH) (right) for a 2D space. The range of each parameter has
been divided into $N_{\rm p}=10$ equally spaced intervals, and each
point has been located so that any column or row does not contain more
than one point. Both designs satisfy the LH condition, but the MLH
design generates a better, less clustered, sampling of the space.}
\label{fig:lh}
\vspace{1cm}
\end{figure}

\subsubsection{Maximum velocity}
In order to avoid particles from reaching arbitrarily high velocities, it is
convenient to limit the maximum velocity they can acquire. For this reason, 
we define the maximum velicity as $V_{\rm max}=0.5(X_{\rm max}-X_{\rm min})$ 
and impose the following restriction
\begin{equation}
v_{k,i}(t) =  
\begin{cases}
 v_{k,{\rm max}} & ~~{\rm when}~~ v_{k,i}(t) > v_{k,{\rm max}} \\ 
-v_{k,{\rm max}} & ~~{\rm when}~~ v_{k,i}(t) < v_{k,{\rm max}} 
\end{cases}	
\end{equation}
where $v_{k,i}(t)$ is the $k$-th component of the velocity for the $i$-th
particle at timestep $t$ and $v_{k,{\rm max}}$ is the $k$-th component of the
maximum velocity defined above. With this condition, the maximum ``jump'' that
a particle can make is equal to half the search space in each dimension.

\subsubsection{Boundary conditions}
We assume reflecting boundary conditions, where the particle reverses the
component of its velocity which is perpendicular to the boundary when it tries
to cross it, i.e.
\begin{equation}
v_{k,i}(t) = -v_{k,i}(t)
\end{equation}
and
\begin{equation}
x_{k,i}(t) = 
\begin{cases}
x_{k,{\rm max}} & ~~{\rm when}~~ x_{k,i}(t) > x_{k,{\rm max}} \\ 
x_{k,{\rm min}} & ~~{\rm when}~~ x_{k,i}(t) < x_{k,{\rm min}} 
\end{cases}
\end{equation}
where $x_{k,i}(t)$ and $v_{k,i}(t)$ are the $k$-th component of the position and
velocity for the $i$-th particle at timestep $t$, respectively, and $x_{k,{\rm
min}}$ and $x_{k,{\rm max}}$ are the $k$-th component of the boundary vectors.

\subsubsection{Convergence criterion}

In all stochastic methods used to explore multidimensional spaces, the
convergence to the region of maximum likelihood is guaranteed only in the
asymptotic limit.  For that reason, any practical implementation of a
stochastic method must include a convergence criterion in order to stop the
exploration, thus assuring that the values of the free parameters have
reached the convergence region.

PSO particles explore the multidimensional space by finding different values of
$G(t)$ as the number of time steps increases. After a certain number of steps,
the value of $G(t)$ becomes stationary and the particles cannot find a best
global value that is significantly different to the current one.  One can take
advantage of this and stop the exploration when the following criterion is
satisfied:
\begin{equation}
\log_{10}(\bar{\Delta}) \le -2.0,
\label{ec:corte}
\end{equation}
where 
\begin{equation}
\bar{\Delta} = \left< {\bar{x}_k(t) - g_k(t) \over g_k(t)} \right>_{k=1,...,{\rm D}},
\end{equation}
and
\begin{equation}
\bar{x}_k(t) = \left< x_{k,i}(t) \right>_{i=1,...,N_{\rm p}},
\end{equation}
where $x_{k,i}(t)$ is the $k$-th component of the $i$-th particle at step $t$,
and $g_k(t)$ is the $k$-th component of $G(t)$ at step $t$.

\subsection{Estimating errors and degeneracies}
\label{sec:errors}

The PSO technique is very efficient to find the region of maximum likelihood,
but it does not provide a detailed description of the neighbourhood of the best
global value, which is needed in order to compute errors and degeneracies. For
this reason, to estimate errors we follow the approach presented by
\cite{prasad_pso_2012}.  The procedure involves taking the subset $j=1,...,M$
of sampled points around the final best global position which satisfy 
\begin{equation}
|\Theta_j| < 0.10~~~{\rm with}~~~\Theta_j = {{X_j(t_{\rm f}) - 
G(t_{\rm f})} \over G(t_{\rm f})},
\end{equation}
where $t_{\rm f}$ is the final step. In this neighbourhood of the best-fitting
point, the fitness function can be approximated by
\begin{equation}
F_j(t_{\rm f}) \simeq F^{\rm max}(\rm t_{\rm f}) \exp{\left(-{1 \over 2}
\Theta_{\it j}^{\rm T}\mathbf{R} \Theta_{\it j} \right)},
\label{eq:gauss}
\end{equation}
where $\mathbf{R}$ is the D$\times$D curvature matrix. The inverse of this
matrix is the covariance matrix $\mathbf{C}$, which can be used to estimate the
error and degeneracies of the parameters around the best global value
\citep{jungman_error_1996}. Taking
\begin{equation}
\Delta_j^2 = -2~{\rm ln}\left[ {F_j(t_{\rm f}) \over F^{\rm max}(t_{\rm f})} 
\right] = \Theta_j^{\rm T} \mathbf{R} \Theta_j,
\end{equation}
a D-dimensional paraboloid can be fitted to the
$\left\{\Delta^2_j\right\}_{j=1,...,M}$ subset to obtain the D(D+1)/2
independent coefficients of the matrix $\mathbf{R}$. The covariance matrix is
then computed taking the inverse of the curvature matrix, i.e.
$\mathbf{C}=\mathbf{R}^{-1}$. The error $\sigma_k$ of the $k$-th parameter is
computed using the diagonal elements of $\mathbf{C}$,
\begin{equation}
\sigma_k = g_k(t)\sqrt{C_{kk}},
\end{equation}
and the off-diagonal elements are used to approximate the possible degeneracies
between parameters.

%
%
%
\section{Constraints for the calibrations}
\label{sec:obs}
\subsection{Observational data}
\label{sec:data}

To calibrate $\sag$ using the PSO method, we compare the properties of the
simulated galaxies with two observationally-determined
statistics focused on the stellar mass content and central SMBH masses
of galaxies, both at $z=0$. These constraints are the stellar mass
function (SMF) and the BH mass to bulge mass relation (BHB):\\

{\bf (1) The stellar mass function at $z=0$ (SMF).}  We calibrate with the data
used by \citet{henriques_mcmc_2014}, which is a combination of the SMF of the Sloan
Digital Sky Survey (SDSS) from \citet{baldry_smf_2008} and \citet{li_smf_2009}, and of the
Galaxy And Mass Assembly (GAMA) from \citet{baldry_smf_2012}.\\ 

{\bf (2) The BH mass to bulge mass relation (BHB).} We combine the datasets
from \cite{mcconnell_bhb_2013} and \cite{kormendy_bhb_2013}.  The data points
used are the average of the BH mass computed over several bulge mass bins, with
errors estimated as the dispersion around that average.\\

The use of SMF to constrain parameters of SAMs is standard practice
\citep{mutch_sam2_2013,mutch_sam_2013,henriques_mcmc_2013,
henriques_mcmc_2014,benson_sam_2014}.
The local SMF is an ideal constraint to evaluate the impact of the physical
processes included in galaxy formation models. In particular,
the low mass end reflects the effect of SNe feedback, whereas
the break and high mass end are directly associated with the impact of AGN
feedback (e.g.  \citealt{benson_sam_2003,bower_agn_2006, lagos_sam_2008}).
Additionally, if a prescription for AGN feedback is implemented (as is the case
in $\sag$), the BHB relationship is a necessary constraint that must also be
satisfied.

\subsection{Likelihood function}

As mentioned in Sec. \ref{sec:method}, for a given position $X$ (i.e., a
particular parameter set) and a given observational property, the likelihood of
the model is computed according to
\begin{equation}
F(X) \propto \exp{\left(-0.5\chi^2\right)},
\end{equation}
with
\begin{equation}
\chi^2 = \sum_{i=1}^{N_{\rm bin}} {{\left( y_{{\rm sag},i} - 
y_{{\rm obs},i}\right)^2} \over {\sigma^2_{{\rm sag},i} + 
\sigma^2_{{\rm obs},i}}}
\end{equation}
where $N_{\rm bin}$ is the number of data bins and $y_{{\rm sag},i}$ and
$y_{{\rm obs},i}$ are the $i$-th values of a given constraint for the {\small
SAG} model and observations, respectively.  The values of $\sigma_{{\rm
sag},i}$ and $\sigma_{{\rm obs},i}$ are the errors for the $i$-th bin for
$\sag$ and the observations, where $\sigma_{{\rm sag},i}$ is computed as
a Poisson error.

The final likelihood assigned to a position $X$ is given by the product of the
likelihoods of all $N_{\rm c}$ constraints used in a particular calibration,
\begin{equation}
F(X) \propto \exp{\left(-0.5\sum_{i=1}^{N_{\rm c}}{\chi^2_i}\right)}
\end{equation}
It is important to note that when we take the direct sum of the $\chi^2$ to
the different constraints to compute the final likelihood, for the sake of
simplicity we are not taking into account the correlation between bins and
observables, which may be non-negligible \citep{benson_sam_2014}.


\section{Results}
\label{sec:results}

\subsection{Best fit values}
\label{sec:bestfit}

\begin{figure}
\centering
\includegraphics[scale=0.70]{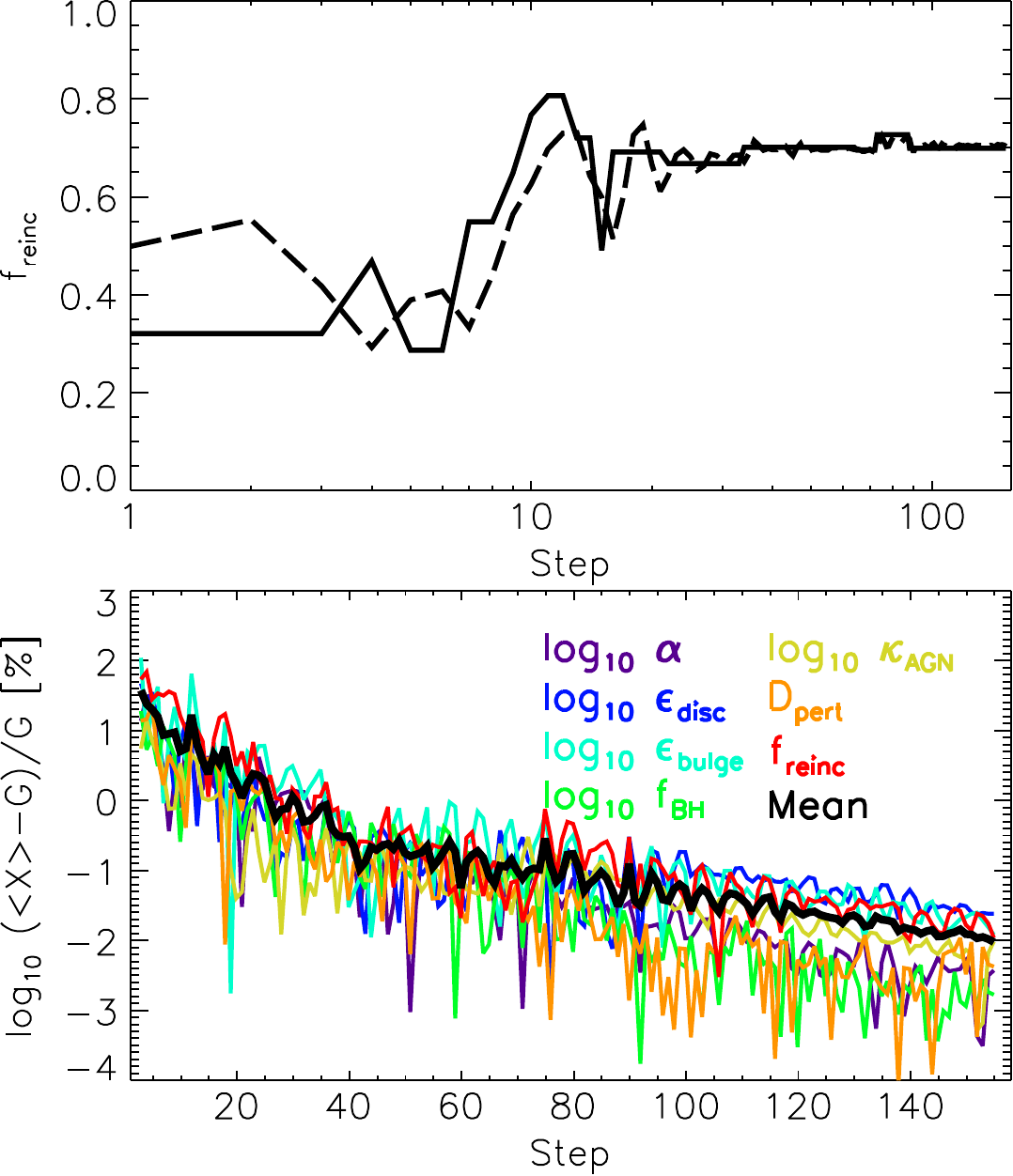}
\caption{
In the top panel we show an example of the temporal evolution of the best
global value $G(t)$ (solid lines) and the average
$\left<X_i(t)\right>_{i=1,...,N_{\rm p}}$ (dashed lines) for the $f_{\rm
reinc}$ parameter. The $y$-axis corresponds to the range explored.
Bottom panel shows the relative difference between the average particle
parameters and the best global value, which is used as convergence criterion.
}
\label{fig:versus} 
\vspace{1cm}
\end{figure}
\begin{figure*}
\centering
\includegraphics[scale=0.55]{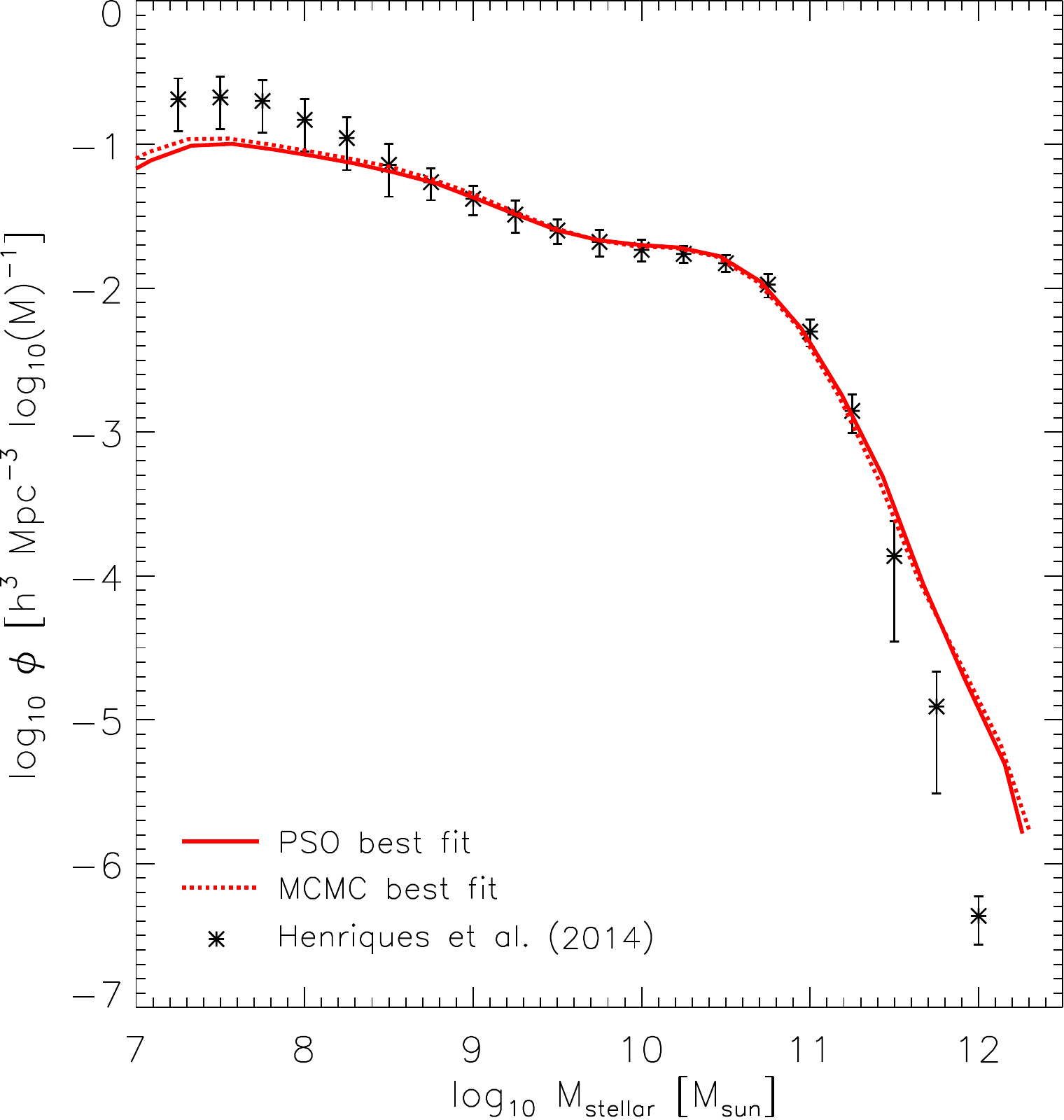}
\includegraphics[scale=0.55]{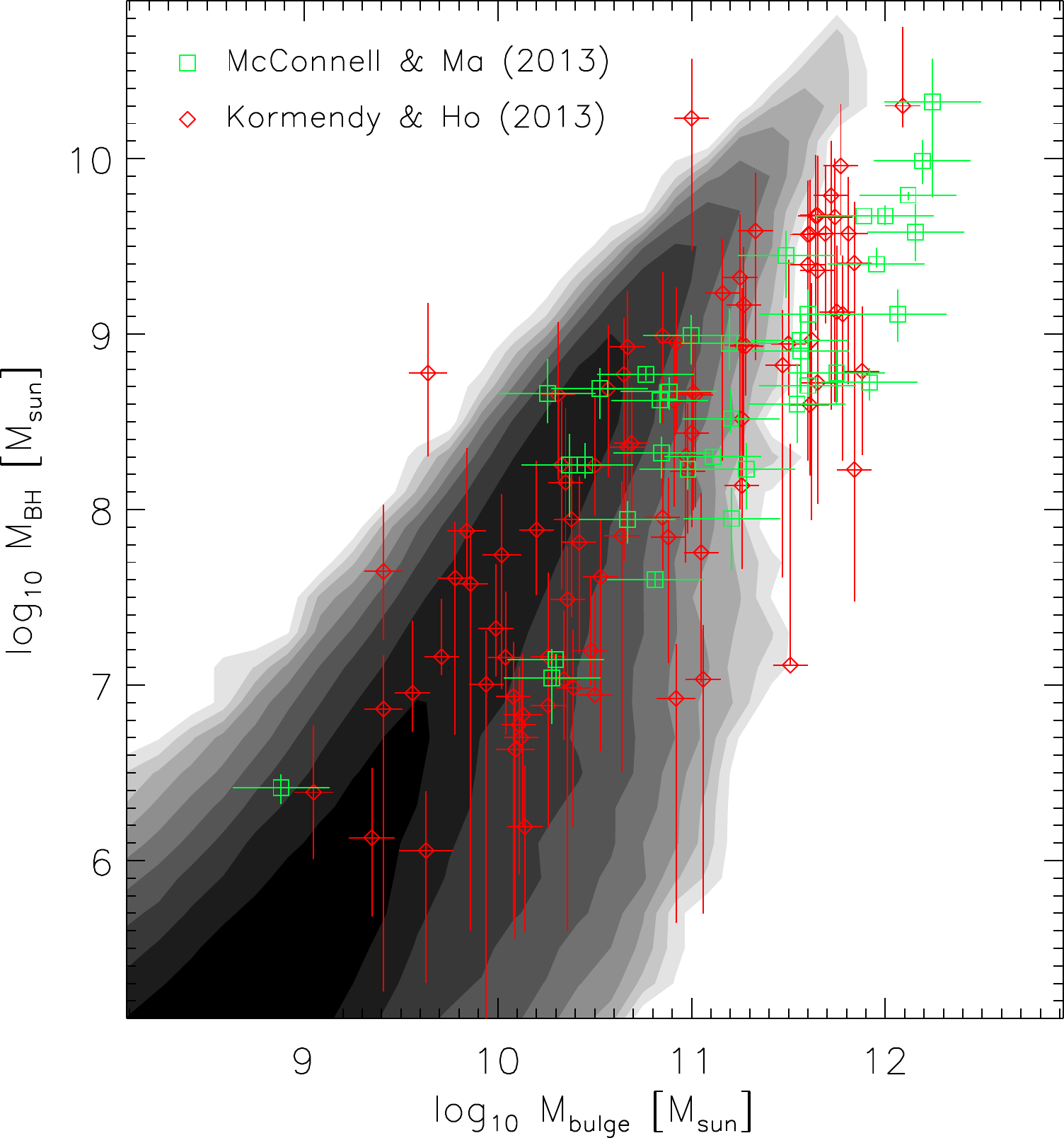}
\caption{
Comparison between the results of the calibrated model and the
observational constraints used. {\em Left panel}: SMF at $z=0$ obtained from
the $\sag$ model calibrated with PSO (red solid line) and with MCMC (red dotted
line); black symbols depict the data compilation of
\citet{henriques_mcmc_2014}.  {\em Right panel}: BHB relation obtained from
$\sag$ galaxies calibrated with PSO (grey contours); results of the calibration
with MCMC are not showed for clarity. The contours represent the normalized
density of the total population of model galaxies that contain a black hole.
Symbols represent the data from \citet{mcconnell_bhb_2013} (green squares) and
\citet{kormendy_bhb_2013} (red rhombus).
}
\label{fig:results}
\vspace{1cm}
\end{figure*}

We consider the observational constraints described in  Sec.
\ref{sec:obs} to calibrate the $\sag$ model. We use $N_{\rm p}=30$ particles
to search for the best fit according to the observational constraints in the
chosen seven-dimensional parameter space. As mentioned in Sec.
\ref{sec:method}, the number of particles $N_{\rm p}$ is a free parameter of
the PSO algorithm and there is no particular criterion to choose a value. In
{\tt PSOS06}, the suggestion for the number of particles of the swarm is
given by $N_{\rm p} = 10 + 2\sqrt{D}$, where $D$ is the dimension of the space.
According to this, we obtain $N_{\rm p} \sim 15$ for our study. We choose to
increase the value to $30$ to have a better sampling of the parameter space and
ensure that similar results --in terms of the best-fitting value-- are obtained
when identical calibrations are performed.

The top panel of Fig.~\ref{fig:versus} shows an example of the evolution of the
best global value $G(t)$ (solid line) and the corresponding average value
$\left<X_i(t)\right>_{i=1,N_{\rm p}}$ (dashed line) of all PSO particles as a
function of the step $t$ for one of the free parameters, $f_{\rm reinc}$.  A
small difference between the global and the average values indicates
convergence to the best set of parameters.  We define the accuracy of
convergence for each parameter in terms of the relative difference between the
average parameter of all the particles and the final best-fitting global value
$G(t)$. The relative difference among all parameters (black line in bottom
panel of Fig.~\ref{fig:versus}) is used as the convergence criterion; the PSO
exploration stops when this difference is smaller than 10$^{-2}$ per cent.  We
can see that the parameters converge in $\sim 150$ steps, which corresponds to
$\sim 4500$ evaluations of the $\sag$ model.

\begin{table}
\caption{Best-fitting parameters found with the PSO and MCMC techniques}.
\centering
\begin{tabular}{ccccc}
\hline                                                                                                                                  
Parameter               & $X_{\rm min}$ & $X_{\rm max}$ & PSO & MCMC        \\
\hline                                                                                                               
$\alpha$                & $0.001$       & $1.0 $        & $0.023\pm0.004$ & $0.022^{+0.003}_{-0.006}$ \\
$\epsilon_{\rm disc}$   & $0.001$       & $1.0 $        & $ 0.19\pm0.01 $ & $ 0.21^{+0.07 }_{-0.05 }$ \\
$\epsilon_{\rm bulge}$  & $0.001$       & $1.0 $        & $ 0.19\pm0.03 $ & $ 0.22^{+0.07 }_{-0.09 }$ \\
$f_{\rm BH}$            & $0.01$        & $1.0 $        & $0.078\pm0.005$ & $ 0.08^{+0.03 }_{-0.03 }$ \\
$10^3 \kappa_{\rm AGN}$ & $0.001$       & $10.0$        & $ 0.08\pm0.01 $ & $ 0.08^{+0.08 }_{-0.04 }$ \\
$D_{\rm pert}$          & $0.0$         & $60.0$        & $   47\pm3    $ & $   49^{+5    }_{-6    }$ \\
$f_{\rm reinc}$         & $0.0$         & $1.0 $        & $ 0.70\pm0.02 $ & $ 0.66^{+0.15 }_{-0.22 }$ \\
\hline                                                                                                                                  
\end{tabular}
\label{tab:bestfit}
\end{table} 

The global best-fitting values found for the calibration done with the PSO
technique are listed in Table~\ref{tab:bestfit} together with the search ranges
$[X_{\rm min},X_{\rm max}]$ and the error estimates obtained according to the
method described in Sec. \ref{sec:errors}.  The reduced chi-square obtained for
the SMF is $\chi^2=0.523$ and for the BHB is $\chi^2=0.208$.

Fig.~\ref{fig:results} shows the results given by the calibrated model for the
statistics involved in the calibration process compared with the corresponding
observational data.  As expected, both observational constraints (SMF at $z=0$
and BHB relation) are well reproduced.  The discrepancies with observational
data at the low and high mass end of the $z=0$ SMF denote some drawbacks in the
model related with the SNe feedback and/or the ejection-reincorporation scheme,
whose effects impact the low mass end, and the AGN feedback that has strong
influence on the high mass end. The latter aspect is also reflected in the BHB
relation obtained from the model, which appears to depart from the observed
data for galaxies with high bulge masses. Since the SMF constraint is stronger
than the one imposed by the BHB relation, the model generates BH with somewhat
larger masses than observed for a galaxy with a given bulge mass in order to
produce a larger AGN feedback that can bring the high mass end of the SMF in
better agreement with observations. However, taking into consideration the high
amount of scatter present in the observational data of the BHB relation, the
calibration can be said to reproduce it satisfactorily.

These results show the good performance of the PSO method when applied to the
calibration process of a SAM.  In order to evaluate the advantages of this
technique with respect to MCMC, we implement the latter method to calibrate our
model using the same set of free parameters and observational constraints. For
this work we run a MCMC of $\sim 30000$ steps using the Metropolis-Hastings
algorithm \citep{metropolis_1953,hastings_1970}, where the tipical length for
SAMs calibrations is between $10^4$ and $10^5$ steps in order to achieve some
level of convergence to the maximum likelihood region
\citep{henriques_mcmc_2009,henriques_mcmc_2013,mutch_sam_2013}. This MCMC
allows us to explore the parameter space with more detail (see Sec.
\ref{sec:space}). The resulting best fit value found by the MCMC is also
presented in Table \ref{tab:bestfit}, with the corresponding $2\sigma$
confidence levels. In this case, the reduced chi-square obtained for the SMF is
$\chi^2=0.513$ and for the BHB is $\chi^2=0.212$. As can be seen, the parameter
values and the final likelihood do not differ significantly from those found
with the PSO algorithm, being statistically consistent.  This fact is reflected
in the left panel of Fig.~\ref{fig:results}, where the SMF at $z=0$ resulting
from the MCMC calibration is also shown.  As can be seen, the SMF predicted by
PSO and MCMC are indistinguishable.
The principal difference between these two results are the errorbars, being
those obtained with MCMC significantly larger.
At this point we need to emphasize that for the PSO method we estimate errors
according to the fitting procedure described in Sec.~\ref{sec:errors},
therefore, the errors in the PSO method must be interpreted just as an
approximation of the true uncertainties.

With this comparison, we demonstrate that the PSO method gives
a definite advantage with respect to MCMC showing convergence to the same
region of maximum likelihood in a considerably smaller computing time.  This is
reaffirmed through the analysis of the parameter space in the next section.

\subsection{The parameter space}
\label{sec:space}

\begin{figure*}
\centering
\includegraphics[scale=1.2]{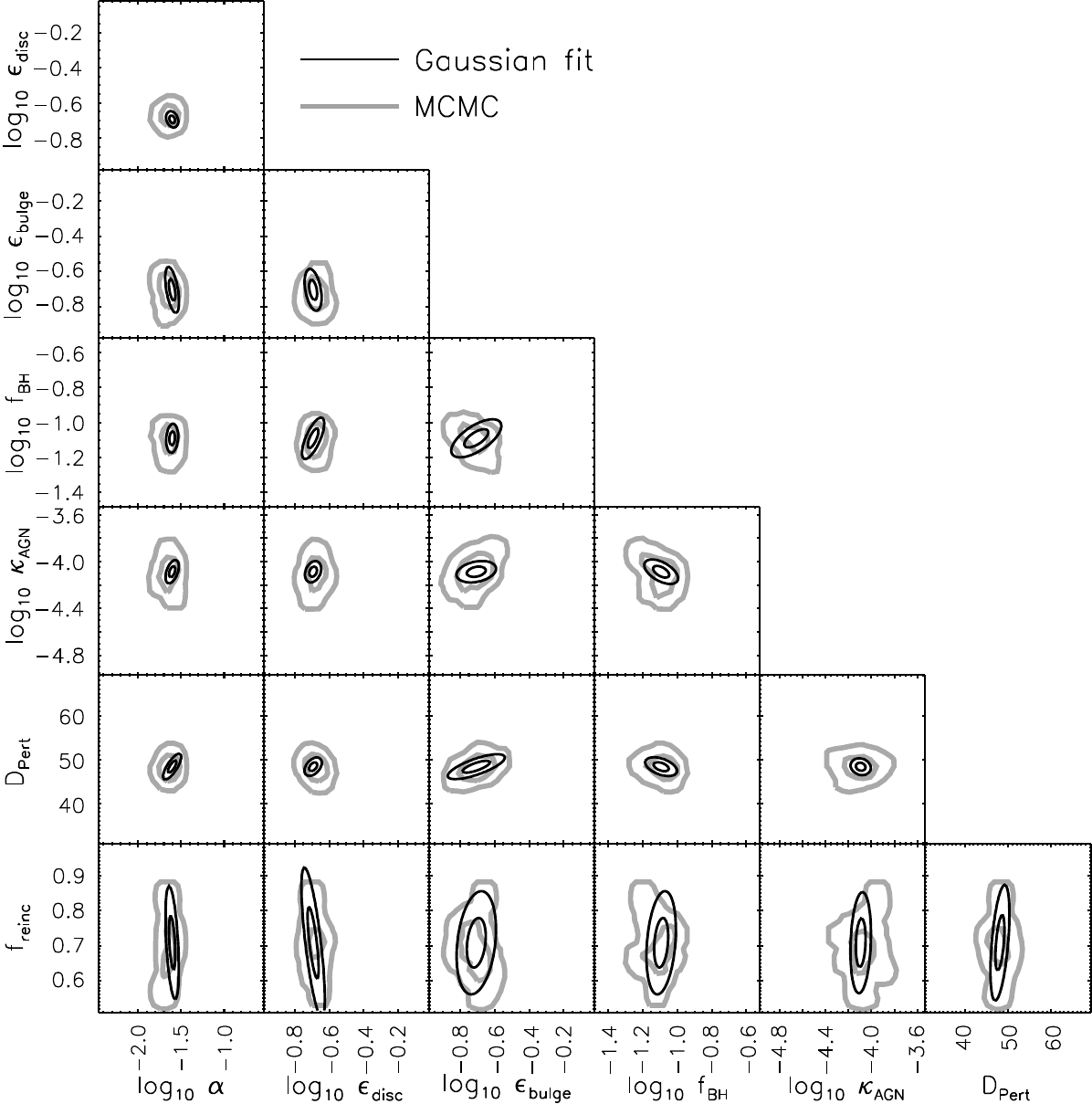}
\caption{
2D projections of the multidimensional Gaussian fit described
in Sec.~\ref{sec:errors} (black thin lines; concentric contours represent the
$0.01$ and $0.5$ levels of the normalised Gaussian), and the $68$ and $95$ per
cent preferred regions of the MCMC (thick grey lines). The projections allow us
to infer information about the degeneracies present between the free parameters
of the model.
}
\vspace{1cm}
\label{fig:space}
\end{figure*}

As mentioned in Sec.~\ref{sec:errors}, the PSO algorithm is extremely efficient
at finding the region of maximum likelihood of a multidimensional parameter
space, but at the expense of not providing a detailed description of the
neighbourhood of this region. However, the fitting procedure presented in
Sec.~\ref{sec:errors} allows us to obtain an estimation of the parameter
degeneracies.

In Fig.~\ref{fig:space} we show the 2D projections of the multidimensional
Gaussian fit (black thin curves) applied in the neighbourhood of the
best-fitting parameters found with PSO. The two concentric contours represent
the $0.01$ and $0.5$ levels of the normalised Gaussian (see
Eq.~\ref{eq:gauss}). In addition, grey thick curves show the $68$ and $95$ per
cent preferred regions of the MCMC exploration performed. It is clear from this
figure that the fitting procedure gives us a decent approximation of the
degeneracies in the parameter space: all the degeneracies present in the
projected Gaussians (with the exception of the $f_{\rm BH}-\epsilon_{\rm
bulge}$ degeneracy) agree with the true ones found with the MCMC sampling. It
is important to recall that the contours of the projected Gaussians do not
represent the same information that the contours of the MCMC (the second is a
true sampling of the parameter space while the first responds to a fitted
curvature matrix in a limited subsample of points of the region of maximum
likelihood), therefore, as we mentioned in Sec.~\ref{sec:bestfit}, this
projected fit must be interpreted just as a merit figure of the true topology
of the parameter space.

The previous considerations demonstrate that the PSO method, combined with the
fitting procedure of Sec.~\ref{sec:errors}, also allows us to infer information
about the parameter degeneracies with no extra evaluations of the model.
However, it is important to emphasize that if an exhaustive exploration of the
parameter space is needed, the PSO method must be used only to locate the
region of the maximum likelihood, and the complementary exploration should be
done using other additional exploration tools, as for instance a localised
MCMC.


\section{Conclusions}
\label{sec:conclusions}

Semi-analytic models of galaxy formation are characterised by a set of free
parameters that regulate the effect of the physical processes involved in
shaping the properties of the galaxy population. We have shown that the PSO
technique can be successfully employed to find the best-fitting set of
parameters for a SAM.  For the particular calibration performed in this work,
the PSO method find the same region of maximum likelihood with 1 order of
magnitude less evaluations than using the traditional MCMC methods, as it has
been already reported by \citet{prasad_pso_2012}.  In this work, we apply the
PSO technique to the $\sag$ galaxy formation model
\citep{cora_sam_2006,lagos_sam_2008,padilla_flips_2014, gargiulo_afe_2014}, but
this method could be applied to any other SAM.

We test the PSO algorithm on $\sag$ by performing a calibration of seven free
parameters of the model using two observational constraints at $z=0$: the
stellar mass function and the black hole to bulge mass relation.  Once the PSO
converges to the best parameter set (i.e., the region of maximum likelihood),
we estimate errors and degeneracies between parameters using the fitting
procedure described in Sec.~\ref{sec:errors}. To validate these results, we
also perform a MCMC exploration, finding the same results not only in terms of
parameter and likelihood values (Fig.~\ref{fig:results} and
Sec.~\ref{sec:bestfit}), but also in terms of the insights of the topology of
the parameter space (Fig.~\ref{fig:space} and Sec.~\ref{sec:space}). The
principal advantage of the PSO algorithm over the traditional MCMC explorations
is the computational cost: while the PSO algorithm needs $\sim 4500$
valuations, the MCMC requires $\sim1$ order of magnitude more. However, it is
fair to remark that the PSO algorithm is a suitable tool to locate ``maximum
values'', whereas the MCMC provides a full exploration method of the parameter
space.

It should be noted that in this work we have only explored parameters related
to the baryonic physics, assuming a fixed cosmological background. However,
this SAM calibration will very likely change if the cosmological parameters
(i.e.  $\Omega_{\rm m}$, $\Omega_{\rm b}$, $\sigma_8$, etc.) are modified.
Recently, useful techniques to scale the cosmology of (sub)halo catalogues
extracted from {\em N}-body simulations were introduced, such as
\citet{angulo_scaling_2010}, \citet{ruiz_scaling_2011} and
\citet{mead_scaling_2014}. Using these techniques, the parameter space of SAMs
can be further extended in order to include both baryonic and cosmological
parameters. As the parameter space of galaxy formation models grows, it becomes
even more complicated to calibrate such models by hand; therefore, calibration
methods such as the one introduced in this paper become a necessity. In further
works we will explore such extended parameter spaces with the new capabilities
provided by the efficient PSO method.

%

\section*{Acknowledgments}

We thank the Referee for his/her careful reading of our manuscript and for
comments and suggestions that help improve this paper.

This work was partially supported by the Consejo Nacional de Investigaciones
Cient\'{\i}ficas y T\'ecnicas (CONICET, Argentina), Secretar\'{\i}a de Ciencia
y Tecnolog\'{\i}a de la Universidad Nacional de C\'ordoba (SeCyT-UNC,
Argentina) and Instituto de Astrof\'isica de La Plata.

The authors kindly acknowledge Volker Springel for providing the
\texttt{SUBFIND} code. We also thank Adam Foster for providing us data of the
total radiated power per chemical element for estimation of the cooling
function.

ANR acknowledges receipt of fellowships from CONICET. SAC acknowledges grants from
CONICET (PIP-220), Agencia Nacional de Promoci\'on Cient\'{\i}fica y
Tecnol\'ogica (PICT-2013-0317), Argentina, and Fondecyt, Chile. TET
acknowledges funding from GEMINI-CONICYT Fund Project 32090021, Comit\'e Mixto
ESO-Chile and Basal-CATA (PFB-06/2007). AO gratefully acknowledges support from
FONDECYT project 3120181. AMMA acknowledges support from CONICYT Doctoral
Fellowship program. 

This project made use of the Geryon cluster at the Centro de Astro-Ingenier\'ia
of Universidad Cat\'olica de Chile for all the calculations performed.

\vspace{1cm}

\bibliographystyle{apj}

\end{document}